# The electronic structure and magnetic phase transition of hexagonal FeSe thin films studied by photoemission spectroscopy

S. Y. Tan[1], C. H. P. Wen[2], M. Xia[2], J. Jiang[2], Q. Song[2], B. P. Xie[2], X. C. Lai[1] and D. L. Feng[2*]

[1] *Science and Technology on Surface Physics and Chemistry Laboratory, Mianyang 621908, China*

[2] *Physics Department, Applied Surface Physics State Key Laboratory, and Advanced Materials Laboratory, Fudan University, Shanghai 200433, China*

Hexagonal FeSe thin films were grown on $SrTiO_3$ substrates and the temperature and thickness dependence of their electronic structures were studied. The hexagonal FeSe is found to be metallic and electron doped, whose Fermi surface consists of six elliptical electron pockets. With decreased temperature, parts of the bands shift downward to high binding energy while some bands shift upwards to $E_F$. The shifts of these bands begin around 300 K and saturate at low temperature, indicating a magnetic phase transition temperature of about 300 K. With increased film thickness, the Fermi surface topology and band structure show no obvious change. Our paper reports the first electronic structure of hexagonal FeSe, and shows that the possible magnetic transition is driven by large scale electronic structure reconstruction.

PACS numbers: 73.20.At, 74.25.Jb, 74.70.Xa, 74.78.Fk

## I. INTRODUCTION

FeSe has the simplest chemical and crystal structure among all the Fe based superconductors. The bulk FeSe superconductor has a superconducting transition temperature (Tc) of ~8 K and it can be dramatically enhanced to 37 K at high pressure[1,2]. Moreover, the discovery of high temperature superconductivity in single unit-cell (UC) FeSe film grown on $SrTiO_3$ (STO) substrate has attracted extensively attention recently[3-9]. No matter for bulk or film materials, the superconductivity has been found to be extreme sensitivity to stoichiometry in $Fe_{1+x}Se$. A superconducting β phase with tetragonal structure and a non-superconducting α phase with hexagonal structure are often found during sample growth with bulk or thin film methods[10-12], which is strongly depending on the Fe/Se ratio and growth temperature. The superconducting phase of tetragonal FeSe has been studied extensively, but little is known about the hexagonal phase, especially for its electronic structure.

Hexagonal structured $FeSe_{1+x}$ (referred to as hexagonal FeSe hereafter) have been found to be magnetic[13,14] and received much attention for its potential application ranging from high-density data storage media[15,16], spintronic devices[17] to anode material for lithium storage[18]. The magnetic and crystallographic properties of hexagonal FeSe strongly depend on the fabrication parameters, crystal structures, and chemical compositions. It is found, within the local-density approximation plus dynamical mean-field theory (LDA+DMFT) that hexagonal FeSe resembles an orbital-selective insulating state[19]. However, hexagonal FeSe usually shows low resistivity in electric measurements with high carrier concentration just like metal[20]. Whether hexagonal FeSe has a metal or a semiconductor nature and the origin of its magnetism are still undetermined.

In this paper, we report an angle-resolved photoemission spectroscopy (ARPES) investigation of the low-energy electronic states of FeSe thin films grown on STO substrates. The hexagonal FeSe is found to be metallic and electron doped, and the Fermi surface is composed of six elliptical electron pockets. The band structure of hexagonal FeSe shows abnormal temperature evolution behavior. With decreased temperature, parts of the bands shift downward to high binding energy while some bands shift upwards to $E_F$. The shifts of these bands begin around 300 K and saturate at low temperature, indicating a magnetic phase transition temperature of about 300 K. With increased film thickness, the Fermi surface topology and band structure show no obvious change except some minor quantum size effect or correlation effect.

## II. EXPERIMENT

High-quality FeSe single crystalline thin films were grown on the $TiO_2$ terminated and Nb-doped (001)-orientated single crystal $SrTiO_3$ (0.5%wt) (Shinkosha) substrate with the molecular beam epitaxy (MBE) method following the previous reports[4,6]. It is reported[11] that NiAs-type hexagonal FeSe can be obtained on double-layer graphene formed on SiC grown at a low substrate temperature of 180 °C. At elevated substrate temperatures exceeds 420 °C, layer-by-layer growth of unstrained epitaxial films of tetragonal β-FeSe can be obtained. Hexagonal and tetragonal structured FeSe can also been grown on tetragonal structured surface of $SrTiO_3$(STO) depending on the substrate temperature. Superconducting tetragonal β-FeSe can be obtained at relatively high substrate temperature of 400-450 °C on $TiO_2$ terminated STO. When the substrate is kept at low temperature of about 300-350 °C, which is just the case of our experiment, we can get the hexagonal FeSe. The heterostructure of FeSe/STO is illustrated in Fig.1(a), and the crystal structure of the obtained NiAs-type hexagonal FeSe film is shown in Fig.1(b).

After growth, the film was directly transferred from the MBE chamber into the ARPES chamber with typical vacuum of $5 \times 10^{-11}$ mbar. ARPES was conducted with 21.2 eV photons from a helium discharge lamp. A SCIENTA R4000 analyzer was used to record ARPES spectra with typical energy and angular resolutions of 10 meV and 0.2°, respectively. A freshly evaporated gold sample in electrical contact with the FeSe sample was served to calibrate $E_F$.

## III. RESULTS AND DISCUSSIONS

The electronic structure of 1 unit-cell (UC) hexagonal FeSe thin film at 30 K is presented in Fig.1. The photoemission intensity map is integrated over a [$E_F$ -10 meV, $E_F$ +10meV] window around the Fermi energy ($E_F$) as shown in Fig.1(e). The observed Fermi surface consists of twelve elliptical electron pockets, which form a beautiful flower-like pattern. The low energy electron diffraction (LEED) pattern of hexagonal FeSe is shown in Fig.1(c), two sets of six-fold symmetric diffraction pattern can be resolved, which indicates that two domain structures exist on the surface of hexagonal FeSe. Based on our ARPES and LEED data, the Fermi surface of single domain hexagonal FeSe is extracted and drawn in Fig.1(d), which contains only six elliptical electron pockets around each M point.

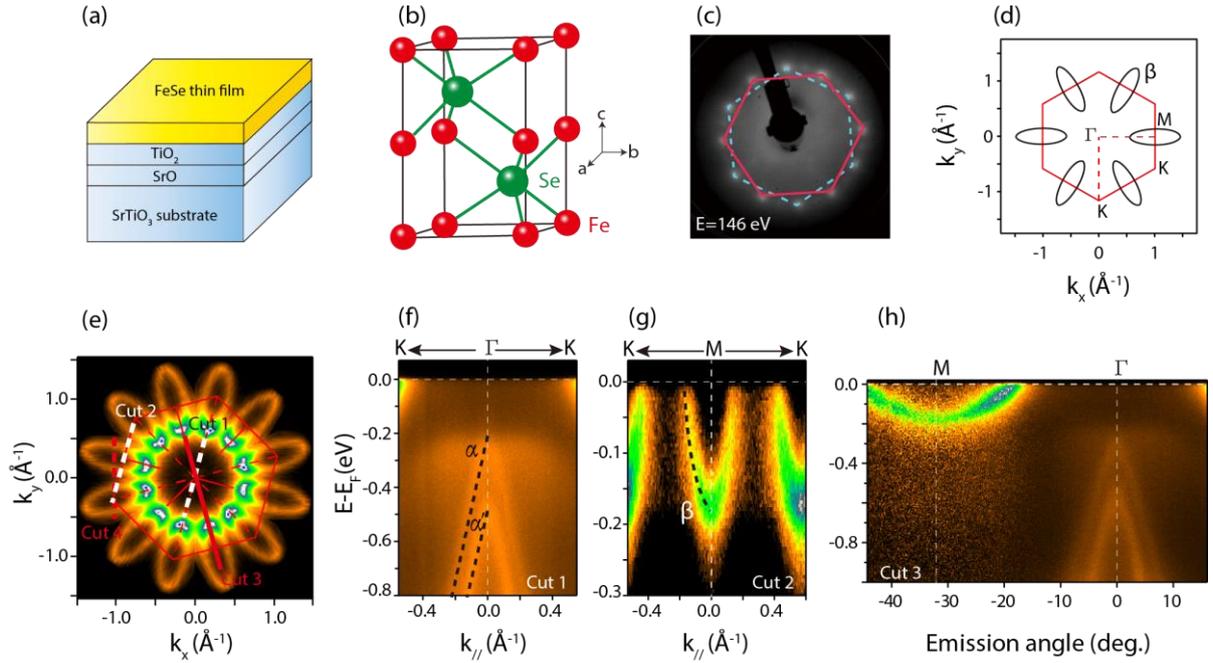

**Fig.1** (Color online) Crystal and electronic structure of 1UC hexagonal FeSe. (a) Schematic structure of the FeSe films on the STO substrate. (b) Hexagonal NiAs-type structure. (c) The LEED pattern of 1UC hexagonal FeSe thin film. (d) The Fermi surface topology extract from (c) and (e). (e) Photoemission intensity map at $E_F$ integrated over [$E_F$ -10 meV, $E_F$ +10 meV]. (f, g) The photoemission intensity plots around Γ and M point, respectively. (h) The photoemission intensity plot along Γ-M direction [cut 3 in (e)]. All the ARPES data were collected at 30 K.

The low-energy band structure along Γ-M direction is shown in Fig.1(h). We can clearly observe an electron-like band (β) centered at M point contributing to the elliptical pockets, whose band bottom located at about $E_F$-0.16 eV[Fig.1(g)]. Two nearly parallel hole-like bands (α and α′) can be seen below $E_F$ at Γ point [Fig.1(f)], whose band top located at $E_F$-0.2 eV and $E_F$-0.47 eV. The existing of only electron band crossing $E_F$ indicates that 1 UC hexagonal FeSe film is electron doped and shows metallic behavior.

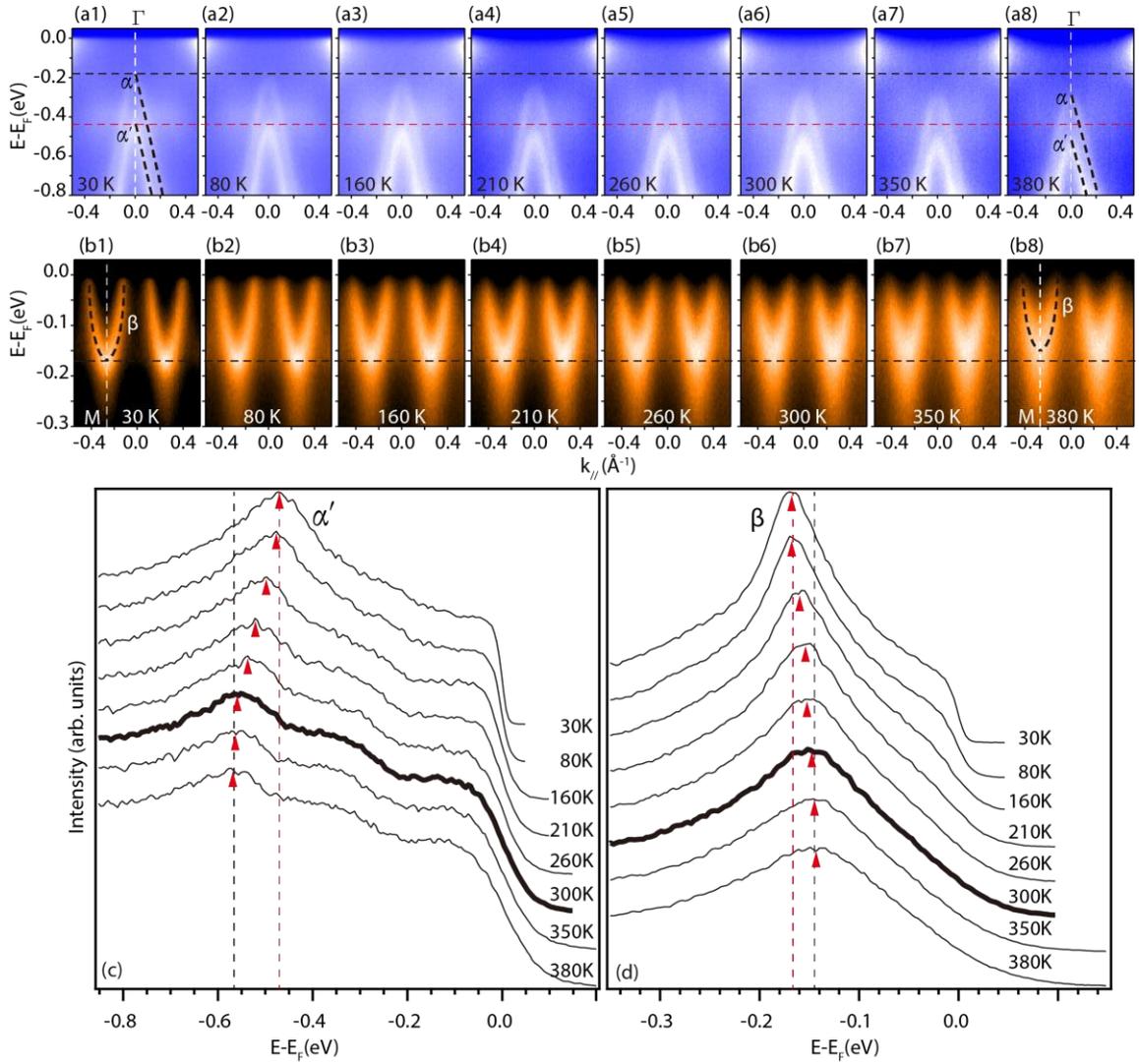

**Fig.2** (Color online) Temperature dependence of the low energy band structure for 1 UC hexagonal FeSe. (a1-a8) Temperature dependence of the band structure around Γ point, [cut 1 in Fig.1(e)]. (b1-b8) Temperature dependence of the band structure around M point, [cut 4 in Fig.1(e)]. (c) The temperature dependence of the energy distribution curves(EDC) at Γ point, the peak position marked by the red triangle correspond to the band top of α′. (d) The temperature dependence of the energy distribution curves(EDC) at M point, the peak position marked by the red triangle correspond to the band bottom of β.

Hexagonal FeSe has always been found to be magnetic(ferrimagnetic or antiferromagnetic depending on the chemical compositions)[13,14]. The band structure of the magnetic materials often get reconstructed across the magnetic transition[21-25]. To examine this, Fig.2 presents the temperature dependence of the low energy band structure of 1 UC hexagonal FeSe. The topologies of the hole (β) and electron (α and α′) bands around Γ and M points show no obvious change, but the positions of these bands shift with changed temperature. The band top of α and α′ gradually shift upward to $E_F$ with decreasing temperature as shown in Fig.2(a), while the band bottom of β gradually shift downward to high binding energy as shown in Fig.2(b).

We tracked the energy distribution curves (EDCs) at the center of α′ and β bands to reveal the temperature dependent band structure evolution more precisely. The parallel α and α′ bands show the same trend of energy shift with changing temperature, we only track the α′ band with much higher intensity as representative. As shown in Fig.2(c), the band top of α′ locates at about $E_F$-0.47 eV at 30 K, which shifts downward with increasing temperature. The band top moved to about $E_F$-0.56 eV at 300 K and kept unchanged even the temperature is further increased, which gives a max energy shift of 90 meV. While for the β band, its band bottom locates at about $E_F$-0.16eV at 30 K, which shifts upward with increasing temperature. The band bottom of β moved to about $E_F$-0.14eV for 300 K and above temperature, which gives a max energy shift of 20 meV.

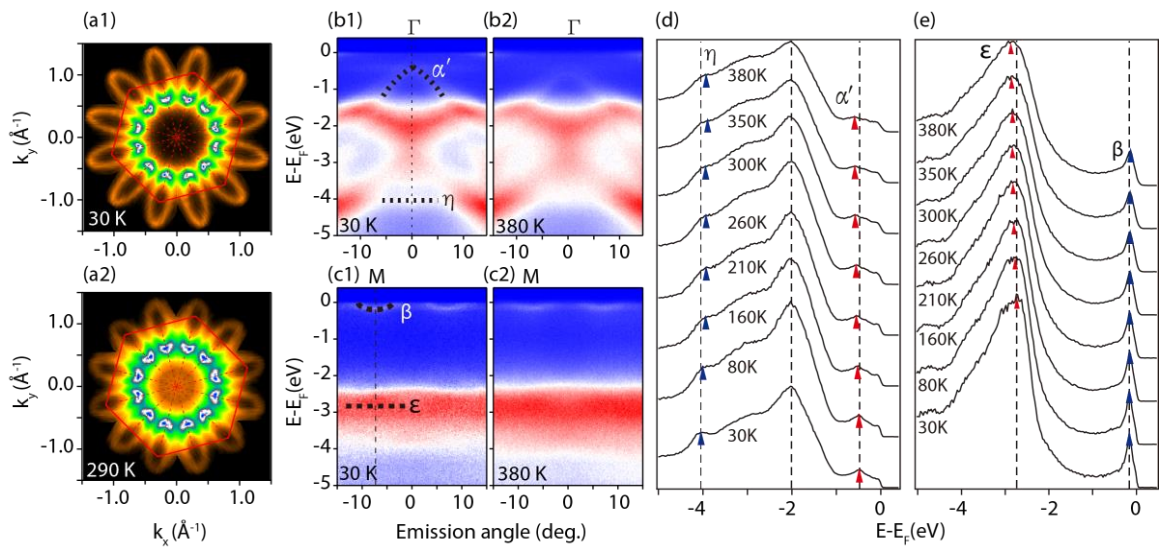

**Fig.3** (Color online) Temperature dependence of the Fermi surface and valence band structure for 1UC

hexagonal FeSe. (a1, a2) The photoemission intensity map at 30 K and 290 K, respectively. (b1, b2) The photoemission intensity plots around Γ point at 30 K and 380 K, respectively. (c1, c2) The photoemission intensity plots around M point at 30 K and 380 K, respectively. (d) The temperature dependence of the energy distribution curves(EDC) at Γ point, the red triangle marked the band top of α′ and the blue triangle marked the band position of η. (d) The temperature dependence of the energy distribution curves(EDC) at M point, the blue triangle marked the band bottom of β and the red triangle marked the band position of ε.

The temperature dependence of the Fermi surface and valence band structure for 1UC hexagonal FeSe are present in Fig.3. The topology of the Fermi surface exhibit negligible change except some thermal broadening at high temperature, and there is no energy gap at the Fermi surface[Figs.3(a1) and (a2)]. To check if there are any other bands shifting with temperature except α/α′ and β, we tracked the EDC around Γ and M points at various temperature as shown in Figs.3d and 3e. Interestingly, the η band at Γ point shifts downward to high binding energy, while ε band around M point shift upward to $E_F$ with decreasing temperature. It is worth noticing that not all the bands shift with temperature, the characteristic peak at about $E_F$-2 eV keeps unchanged at various temperatures[Fig.3(d)]. To summarize, with decreased temperature, the band shift is +90 meV and +80 meV upward for α/α′ and ε, -20meV and -15 meV downward for β and η. We note that the upward shifts are larger than the downward shifts with decreasing temperature, and it seems that the electronic energy is not reduced in the possible magnetic ground state. Presumably, the bands would shift towards higher binding energy to reduce energy, which is inconsistent with the actual situation, further studies are needed to fully understand this issue.

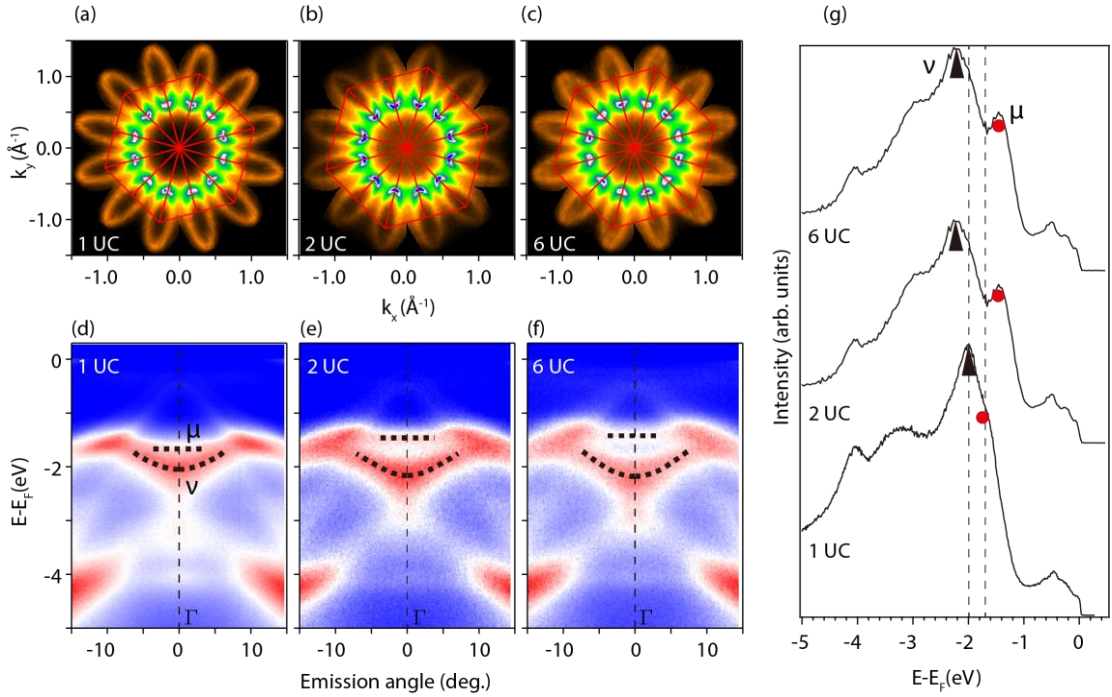

**Fig.4** (Color online) Thickness dependence of the Fermi surface and valence band structure for hexagonal FeSe at 30 K. (a, b, c) The photoemission intensity map of 1 UC, 2 UC and 6 UC hexagonal FeSe, respectively. (d, e, f) The photoemission intensity plots around Γ point of 1 UC, 2 UC and 6 UC hexagonal FeSe, respectively. (g) Thickness dependence of the energy distribution curves (EDC) at Γ point, the black triangle marked the band bottom of ν and the red circle marked the band position of μ.

The Fermi surface and band structure of tetragonal FeSe thin films have been found to change dramatically with different thickness[6,9]. It is natural to ask whether the electronic structure of hexagonal FeSe has similar thickness dependence? The electronic structure of 1 UC, 2 UC and 6 UC hexagonal FeSe are shown in Fig.4. Due to the mismatch of the lattice symmetry between hexagonal FeSe and tetragonal STO substrate, the sample quality become worse and worse with increased film thickness, and clear band dispersion can't be detected in thick thin films. The obtained Fermi surface topology and valence band structure show no dramatic change for different film thickness. However, subtle changes of some typical bands still exist in the valence band structure around Γ point. The position of μ and ν bands shift with changing film thickness, as marked and shown in Figs.4(d-f). The band bottom of ν shifts downward with increasing film thickness, while the μ band shifts upward as shown in Fig.4g. The intrinsic nature of this thickness dependent band shift is not clear by now, which may be associated with the quantum size effect[26,27] or the reducing correlation with increased film thickness.

## IV. DISCUSSION AND CONCLUSION

Hexagonal $FeSe_{1+x}$ exhibits both antiferromagnetism and ferrimagnetism depending on composition. According to the magnetic phase diagram[14] of hexagonal $FeSe_{1+x}$ reported by P.Terzieff, $FeSe_{1+x}$ shows antiferromagnetism with compositions of $0.02 \leq x \leq 0.10$ (50.5 to 52.5 at% Se), ferrimagnetism is observed at compositions near $0.10 \leq x \leq 0.36$ (52.5 to 61.0 at% Se), and both the Neel and Curie temperatures vary with composition. The precise composition of our thin film samples can't be determined due to the lack of in-situ characterization method, but the low growth/annealing temperature used in our study undoubtedly lead to a bit Se rich $FeSe_{1+x}$ samples[11], which is most likely antiferromagnetic in our case.

The band shift behaviors observed in our hexagonal FeSe thin films are not exclusive in magnetic compounds. Similar band shift behavior has been reported in $TaFe_{1.23}Te_3$, a spin ladder compound with antiferromagnetic ground state[28]. It is proposed that $TaFe_{1.23}Te_3$ is the second kind of novel quantum material in addition to the parent compounds of iron-based superconductors, whose AFM transition directly correlates with the electronic structure reconstruction at high binding energies. With a simple chemical and crystal structure, clear experimental band dispersion, the hexagonal FeSe provides an excellent platform for experimental and theoretical study of the mechanism for magnetic transition.

In summary, we report the first in situ angle-resolved photoemission spectroscopy study of hexagonal FeSe thin films. The hexagonal FeSe is found to be metallic and electron doped, and the Fermi surface consists of six elliptic electron pockets. With decreased temperature, parts of the bands shift downward to high binding energy while some bands shift upwards to $E_F$. The shifts of these bands begin around 300 K, and saturate at low temperature, indicating a magnetic phase transition temperature of about 300 K. With increased film thickness, the Fermi surface topology and band structure show no obvious change. Our paper shows that the possible magnetic transition in hexagonal FeSe is driven by electronic structure reconstruction, which is crucial to understand the origin of the magnetism in NiAs-type FeSe.

## Acknowledgements

We gratefully acknowledge helpful discussions with Prof. X. G. Gong, X. Dai, and Dr. H. Y. Cao. This work is supported by the Foundation of President of China Academy of Engineering Physics (Grants No. 201501037).

---

*dlfeng@fudan.edu.cn